%% file: main.tex
\documentclass[%
 reprint,
 english,
 hyphens,
superscriptaddress,
 amsmath,amssymb,
 aps,
floatfix,
]{revtex4-2}

\usepackage{graphicx}% Include figure files
\usepackage{dcolumn}% Align table columns on decimal point
\usepackage{bm}% bold math
\usepackage{hyperref}% add hypertext capabilities
\usepackage{float}
\usepackage{braket}
\usepackage{booktabs}
\usepackage{makecell}
\usepackage[dvipsnames]{xcolor}

\usepackage{natbib}
\usepackage{bibentry} % this is to put full citations in "Compare LAC Interpretations"
\usepackage{dblfloatfix}
\usepackage{amsmath}
\usepackage{subfiles} % Best loaded last in the preamble
\usepackage[english]{babel}
\usepackage{placeins}
\makeatletter
\addto\extrasenglish{%
  \expandafter\let\csname captionsen\endcsname\captionsenglish
  \expandafter\let\csname dateen\endcsname\dateenglish
}
\makeatother
\makeatletter
\renewcommand{\selectlanguage}[1]{\relax}
\makeatother

\begin{document}

\title{Achieving High Filling of an Optical Lattice by Light-Assisted Redistribution of Atoms}

\author{Lauren Weiss}
\email{laurensweiss@gmail.com}  
\affiliation{James Franck Institute, Enrico Fermi Institute, and Department of Physics, University of Chicago, Chicago, Illinois 60637, USA}
\author{Evan Yamaguchi}
\affiliation{James Franck Institute, Enrico Fermi Institute, and Department of Physics, University of Chicago, Chicago, Illinois 60637, USA}
\affiliation{Pritzker School of Molecular Engineering, University of Chicago, Chicago, Illinois 60637, USA}
\author{Claire Pritts}
\affiliation{James Franck Institute, Enrico Fermi Institute, and Department of Physics, University of Chicago, Chicago, Illinois 60637, USA}
\author{Tadej Me{\v z}nar{\v s}i{\v c}}
\affiliation{James Franck Institute, Enrico Fermi Institute, and Department of Physics, University of Chicago, Chicago, Illinois 60637, USA}
\author{Cheng Chin}
\affiliation{James Franck Institute, Enrico Fermi Institute, and Department of Physics, University of Chicago, Chicago, Illinois 60637, USA}

\begin{abstract} 

Scalable arrays of individual atoms provide an ideal starting point for quantum information and simulation experiments. However, their preparation is often limited by light-assisted collisions (LACs), which typically result in parity-projected filling fractions of $f \approx 0.5$. In this work, we demonstrate a light-assisted redistribution process in the Quantum Matter Synthesizer that overcomes this constraint by stochastically moving atoms from multiply occupied lattice sites to neighboring vacant sites. Using a blue-detuned optical pumping beam during degenerate Raman sideband cooling, we achieve single-atom filling fractions of $70-80\%$. We find that over 50$\%$ of the atoms involved in radiative collisions are retained in the lattice. The redistribution process involves many LACs over an extended time as atoms diffuse to empty sites. Our demonstration offers a scalable and efficient pathway toward unity-filled atom arrays without the need for complex rearrangement protocols, with broad applicability to quantum simulation, precision measurements, and quantum information control.
\end{abstract}

\maketitle
\section{Introduction}
Arrays of single atoms have emerged as a premier platform for quantum science research \cite{singh_dual-element_2022,barnes_assembly_2022, pichard_rearrangement_2024,barredo_atom-by-atom_2016,manetsch_tweezer_2025,bluvstein_logical_2024} and quantum simulation \cite{sherson_single-atom-resolved_2010, parsons_site-resolved_2015}. A central challenge in preparing such systems is the effect of light-assisted collisions (LACs). In the presence of near-resonant light, two or more atoms occupying a single trap are frequently excited to molecular potentials and subsequently lost in pairs \cite{pampel_quantifying_2025_1, fuhrmanek_study_2012}. This process leads to parity-projected imaging in quantum gas microscope experiments \cite{depue_unity_1999,bakr_quantum-gas_2009,buob_strontium_2024,endres_single-site-_2013}, and also limits the atom filling to $f\approx0.5-0.6$ in many optical tweezer experiments. Sophisticated schemes are required to avoid parity-projection and detect multiple atoms in a site \cite{su_fast_2025, falcon_microsecond-scale_2025}.

One method to reach near-perfect unity-filled optical lattices is to load an atomic Bose-Einstein condensate into a lattice and convert it into the $n=1$ Mott insulator \cite{jaksch_cold_1998, greiner_quantum_2002,bakr_probing_2010, tomita_observation_2017}. While this method requires longer evaporative cooling, a filling fraction of over 90$\%$ can be achieved. In optical tweezer experiments where tweezers can be quickly loaded with a thermal sample, rearrangement of individual tweezers is required to assemble unity-filled arrays \cite{eckner_assembling_2025,gyger_continuous_2024,endres_atom-by-atom_2016}, adding significant experimental complexity.

\begin{figure}[h]
\includegraphics[width=1\linewidth]{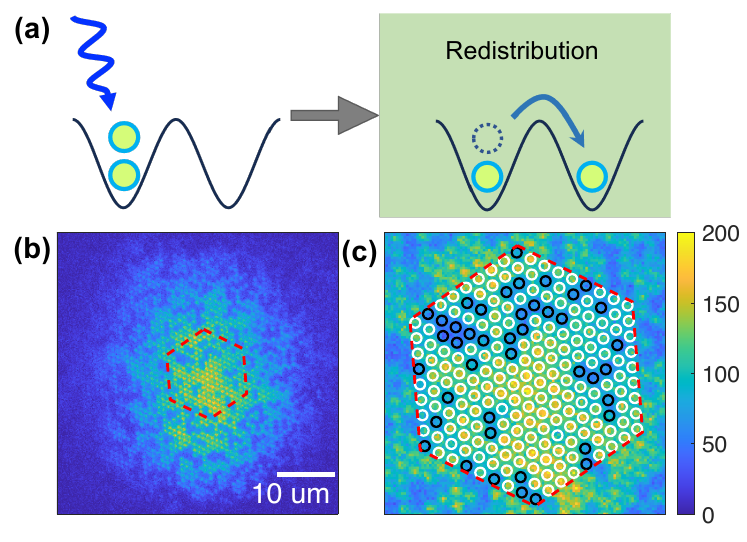} 
   \caption{High filling of atoms in optical lattices by light-assisted redistribution. (a) Redistribution of atoms from a multiply occupied site to a vacant site enhances unity filling fidelity. (b) Site-resolved image of thermal atoms in the lattices with high filling fraction $f$ = 0.80 within the central 217-site region (red dashed line). (c) Enlarged view of the region of interest. Lattice constant is $a = 880$~nm.}
   \label{fig:interactions}
\end{figure}

Scalable schemes have been realized to achieve higher preparation fidelity in optical tweezers. Using a blue-detuned laser beam, it is possible to induce LACs to repel and eject atoms from a trap until one remains \cite{grunzweig_near-deterministic_2010}. This light-induced repulsion can prepare unity filling of a few alkali atoms in tweezers with over 90$\%$ fidelity \cite{carpentier_preparation_2013, brown_gray-molasses_2019} and about 80\% loading probability in an array of over 100 tweezers \cite{brown_gray-molasses_2019}. Experiments with alkaline-earth atoms have achieved even higher loading fraction, $>96\%$ for a single tweezer and $>92\%$ for an array of 100 tweezers \cite{jenkins_ytterbium_2022}, and $83.5\%$ loading for an array of 2400 tweezers \cite{zhu_high-efficiency_2025}. %These results suggest that blue-detuned LACs can also enable high filling fractions for thermal atoms in optical lattices. 

In this work, we demonstrate a light-assisted redistribution process in an optical lattice, which disperses atoms from multiply occupied sites into vacant sites, resulting in a high filling fraction (Fig~\ref{fig:interactions}a). \textcolor{black}{When multiple atoms in a lattice site are subject to blue-detuned LACs, atoms can be ejected from the site. Similar to optical tweezer experiments ~\cite{lester_rapid_2015}, we observe that some of these singly-ejected atoms can be recaptured in neighboring sites.} This redistribution process allows us to reach filling fractions of $70\sim80~\%$, (Fig~\ref{fig:interactions}b). We provide direct spatial evidence of this mechanism by observing atoms migrating into initially unoccupied lattice sites. By analyzing the filling and loss dynamics, we show that over 50$\%$ of atoms participating in repeated radiative collisions ultimately remain trapped in the lattice.

\section{Experimental procedure}
We perform the experiment in the Quantum Matter Synthesizer apparatus, which combines an optical lattice, optical tweezers and quantum gas microscope to prepare, control and detect individual atoms \cite{trisnadi_design_2022,supplement}. Our experiment starts with precooled Cs atoms transferred from the magneto-optical trap to a glass cell, where the atoms are loaded into a two-dimensional (2D) triangular optical lattice in the horizontal plane. The lattice is formed at the center of the cell by intersecting three circularly polarized 935~nm beams \cite{trisnadi_design_2022}. In the 2D lattice, we perform degenerate Raman sideband cooling (dRSC) \cite{supplement,vuletic_degenerate_1998,kerman_beyond_2000,han_3d_2000} on the D2 line \cite{qiu_loading_2025}. We then release the atoms into a 1064~nm light sheet dipole trap. After thermalization in the light sheet, we obtain a thermal sample of 2000 atoms at temperature $T\approx 15~\mu$K. 

\textcolor{black}{The atoms are then loaded back to the 2D lattice by adiabatically ramping the lattice to its full depth of 150 $\mu$K. } Within the central 217-site region of interest, we estimate an average site occupancy of $\bar{n}=2$ atoms per lattice site. \textcolor{black}{The average site occupancy depends on the number of atoms loaded into the lattice, which is controlled by the magneto-optical trap (MOT) loading time.} We then perform another stage of dRSC on the atoms in the presence of the light sheet. In this stage of cooling, which we call the LAC-cooling stage, we tune the intensity $I$ and the frequency of the optical pumping (OP) beam to investigate LACs between atoms in multiply-occupied sites. \textcolor{black}{
On the other hand, atoms in singly-occupied sites are not subject to LACs and remain cold and trapped.}

To detect the atomic distribution in the lattice, we perform fluorescence imaging on the atoms. The imaging routine employs the dRSC with a strong OP beam for 200~ms that scatters sufficient photons to the camera. The imaging also causes parity projection on the atom occupancy as excess atoms are lost in pairs; we observe zero or one atom per site in the images \cite{supplement}. We define the filling fraction $f$ as the fraction of occupied sites within the region of interest. Fig.~\ref{fig:interactions}(b) and (c) show an example image with optimal filling of 80\% in the region of interest. We can also perform absorption imaging on the sample after a short time-of-flight expansion. This method is not subject to parity projection and allows us to measure the total atom number $N_{\text{tot}}$, including those in multiply-occupied sites.

\begin{figure}[t]
\centering
\includegraphics[width=\linewidth]{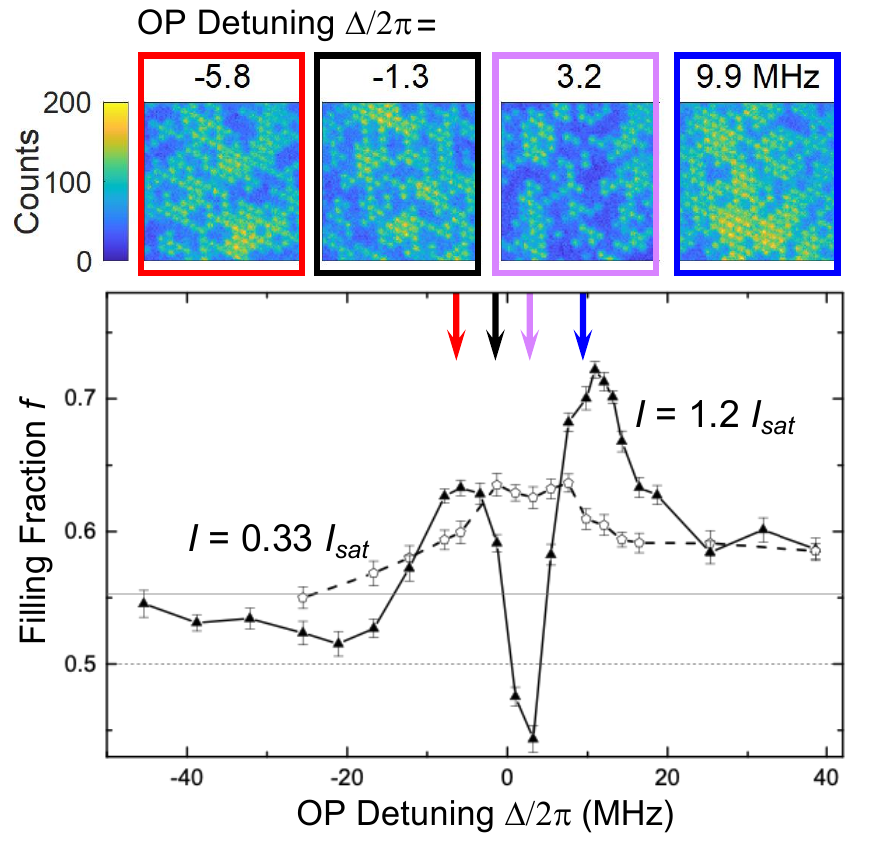} 
   \caption{Control of filling fraction by optical pumping. In the LAC-cooling stage, an OP beam is applied to the atoms in lattices for 125~ms. The filling fractions $f$ are calculated from 20 repeated experiments within the 217-site central region. The OP beam is detuned relative to the $\ket{6S,F=3} \rightarrow \ket{6P_{3/2},F'=2}$ transition by $\Delta$. The OP intensities are $I= 0.33I_{\mathrm{sat}}$ (open circles) and $1.2I_{\mathrm{sat}}$ (filled circles), where we obtain the optimum filling at $\Delta/2\pi=$+9.9 MHz. The data are compared with the filling fraction without the LAC-cooling $f = 0.55$ (gray line) and the theoretical upper limit of filling fraction for a thermal gas with parity projection $f = 0.5$ (dashed line). Error bars show standard deviations of approximately 3~\%. Upper images show examples with OP intensity $I=1.2I_{\mathrm{sat}}$.}
      \label{fig:freq}
\end{figure}

\section{Results and discussion}
The filling fraction varies drastically depending on the OP parameters during the LAC-cooling stage (Fig.~\ref{fig:freq}). At low OP intensity $I < I_{\text{sat}}$, where $I_{\text{sat}}=1.1~$mW/cm$^2$ is the Cs saturation intensity, the filling fraction exhibits a modest and symmetric enhancement on both sides of the atomic resonance. At higher intensities, the filling develops two well-separated peaks on the red- and blue-detuned sides and a dip near resonance. A maximum filling fraction exceeding 70~$\%$ is achieved for blue detuning $\Delta/2\pi=10$~MHz and intensity $I=1.2 I_{\text{sat}}$. For even higher intensities, the filling declines. 

The behavior at low OP intensity $I<I_{\text{sat}}$ can be described by reabsorption \cite{khaykovich_saturation_1999}, which occurs when a photon scattered by an atom is absorbed by a nearby atom. Theory suggests enhanced reabsorption symmetrically on both sides of the bare atomic resonance at the detuning given by the Rabi frequency \cite{khaykovich_saturation_1999}, fully consistent with our observation. Reabsorption leads to simultaneous excitation of multiple atoms within the same site \cite{castin_reabsorption_1998}, and can repeatedly heat atoms until only one remains in the lattice site. Furthermore, escaping atoms can be recaptured and cooled into vacant sites, enhancing the filling \cite{gisbert_stochastic_2019,kerman_beyond_2000}.

The maximum filling fraction occurs at a high OP intensity $I =1.2 I_{\text{sat}}$. Clear asymmetry manifests in the filling fraction with respect to the OP detuning with higher filling on the blue-detuned side $\Delta > 0$. No such asymmetry is observed in the cooling performance of isolated atoms. We attribute this asymmetry to the coupling of colliding atoms to the repulsive (attractive) molecular potential for positive (negative) detuning. In our experiment, the blue-detuned OP beam efficiently excites and repels scattering atoms to neighboring sites to enhance filling. Red-detuned light, on the other hand, causes excited atoms to attract each other, leading to a higher probability to form molecules, which are lost from the trap. On resonance $\Delta = 0$, the filling drops below the thermal gas limit $f < 0.5$, which we attribute to the reduced cooling performance \cite{vuletic_degenerate_1998}.

\begin{figure}[t]
\includegraphics[width=1\linewidth]{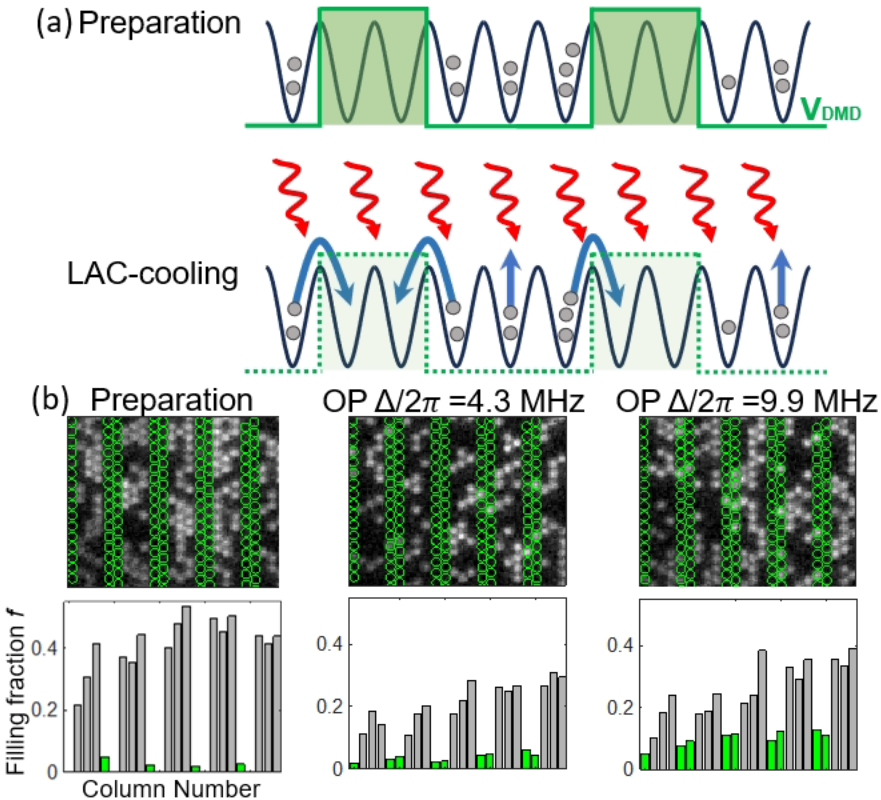}
\caption{Evidence of light-assisted redistribution of atoms. (a) In preparation, atoms are depleted from two out of every five columns of the lattice sites with a digital micromirror device (DMD), indicated by the shaded green area. In LAC-cooling stage, OP beam (red waves) induces redistribution and radiative loss (blue arrow) in sites with multiple atoms. The final atomic distribution is detected with site-resolved imaging. (b) Single-shot images of atoms from the preparation (left), LAC-cooling stages with near-resonant OP (middle), and blue-detuned OP (right). Green circles indicate sites emptied in the preparation. Lower panel shows the column-averaged occupancies from 20 experiments.}
\label{fig:green}
\end{figure}

\subsection{Light-assisted redistribution}
To confirm the atomic redistribution, we deliberately pattern atoms in the lattice and study their distribution after the LAC-cooling stage. After loading the atoms into the 2D lattice, we project a fast-switching optical potential to remove atoms from selected columns of lattice sites \cite{supplement}. This procedure efficiently removes atoms from the illuminated sites, leading to a low atom occupancy of $1.4(7) \%$ while the unilluminated sites have a mean atom occupancy of $f=40(1)\%$ (Fig~\ref{fig:green}). After applying LAC-cooling for 70~ms we perform site-resolved imaging on the atoms. We find that a sizable fraction of the atoms are transferred to the emptied sites during the LAC-cooling stage, which provides clear evidence that the OP beam redistributes the atoms in the lattice (Fig.~\ref{fig:green}b). For blue-detuned OP ($\Delta/2\pi = 9.9$ MHz) in particular, the emptied sites are filled up to 10(1)\% compared to 3.7(4)$\%$ with near-resonant OP \cite{supplement}. \textcolor{black}{Once atoms are in singly occupied sites, they are free from LACs and fully localized in the lattice sites.}

To extract the fraction of atoms that are redistributed during the LAC-cooling stage, we introduce a thermal model to describe the initial distribution of atoms in the lattice with Poisson statistics. The measured filling fraction $f = 0.4$ of the unilluminated sites suggests an average site occupancy of $\lambda = -\frac{1}{2}\ln(1-2f)=0.8$ before LAC-cooling. This implies that, on average, $f_r = \lambda(1-e^{-\lambda})=0.44$ atoms per site participate in LAC processes. We assume these excited atoms are redistributed by LACs to neighboring sites with equal probability. If an atom moves to an empty site, LACs stop, and the atom will be cooled and settle in the site. If the atom reaches an occupied site, further LACs or collisional blockade  will repel atoms to another neighboring site \cite{fung_efficient_2015, fuhrmanek_study_2012, schlosser_collisional_2002,forster_number-triggered_2006}. From the filling increase in the illuminated columns, $\Delta f = 8.8\%$ for blue-detuned OP, we estimate that a large fraction $\frac{6\Delta f}{2f_r}=60(6) \%$ of the atoms in a multiply-occupied site redistribute to its six neighbors. 

\subsection{Dynamics of light-assisted redistribution}
\begin{figure}[t]
\centering
\includegraphics[width=0.99\linewidth]{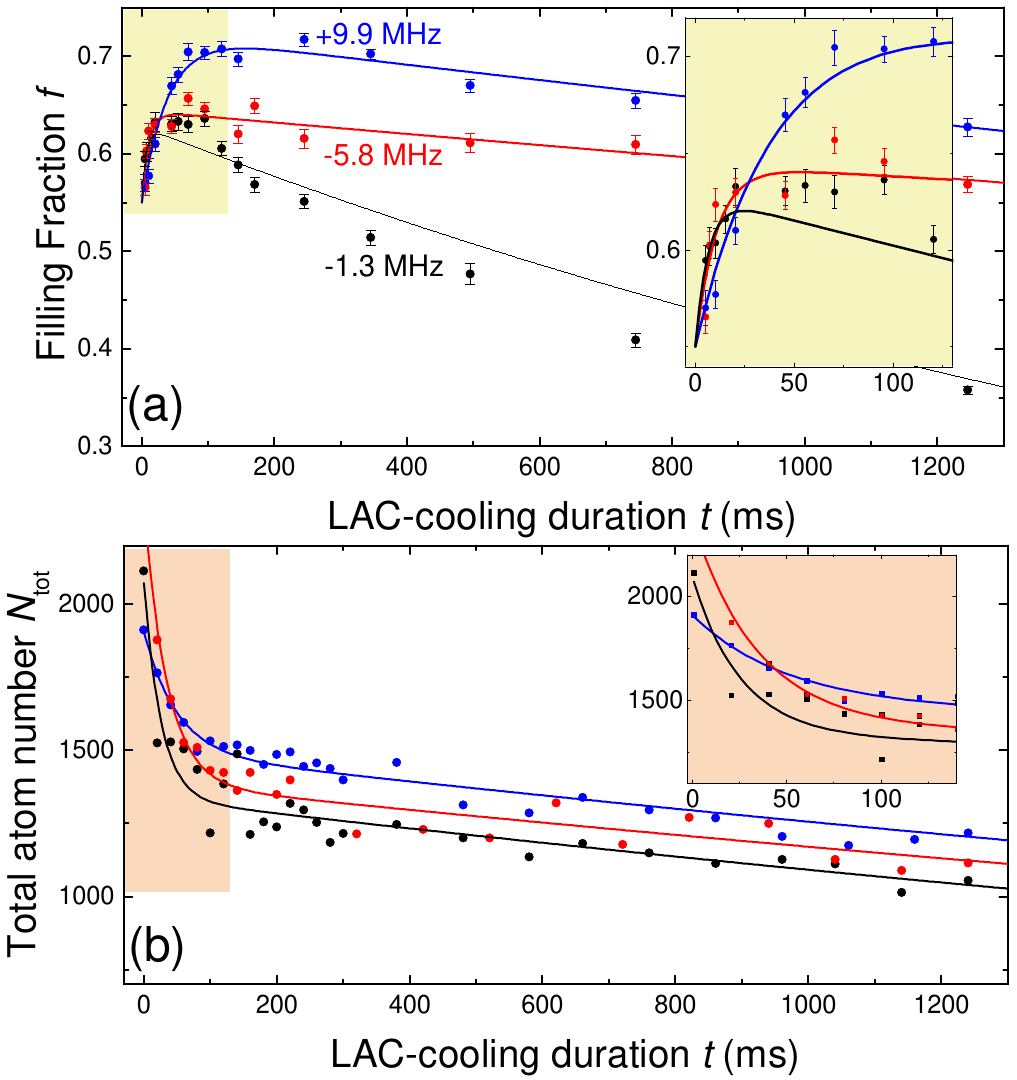} 
   \caption{Time evolution of light-assisted enhancement of filling fractions in  optical lattice. Optical pumping detuning is $\Delta/2\pi =$ -5.8 MHz (red), -1.3 MHz (black) and +9.9 MHz (blue) from the $|6S_{1/2}, F=3\rangle$  to $|6P_{3/2},F'=2\rangle$ transition. (a) Filling fraction $f$ in the central 217-site based on 20 site-resolved images. (b) Total atom number $N_{\text{tot}}$ from absorption images. Lines are double-exponential fits based on our model \cite{supplement}. Insets show the early time behavior.}
   \label{fig:time}
\end{figure}

To understand the dynamics of the light-induced redistribution, we apply the LAC-cooling on a dense thermal sample in the lattice for various interaction times $t$. We then measure the atomic distribution from site-resolved imaging and the total atom number $N_{\text{tot}}$ from time-of-flight measurement (Fig.~\ref{fig:time}a, b).  The loading dynamics in tweezer arrays have also been investigated \cite{grunzweig_near-deterministic_2010, brown_gray-molasses_2019}.

In early times $t<100$~ms, we observe rapid increase of filling fraction and the atom number decays by $25$--$40\%$. Both processes are characterized by similar short timescales, which indicates they are associated with the same LAC process. For the blue-detuned OP with $\Delta/2\pi=+9.9$ MHz, which yields the highest filling fraction $f>0.7$, the dynamics are much slower than for red-detuned and near resonant beam, where we observe a smaller gain in the filling and greater initial atom loss (Fig.~\ref{fig:time}).

At later times $t>200$~ms, all atoms settle to singly-occupied sites and thus radiative collisions stop. Here the atom number shows a slow exponential decay with a lifetime of $5.8(3)$~s, limited by collisions with the background gas. The filling fraction decay time depends on the OP frequencies due to the variation of the dRSC efficiencies in the regime of interest. When counting all atoms in the site-resolved images, $N_{\text{site}}$, we obtain the decay time of $6(2)$~s \cite{supplement}. The excellent agreement between the two imaging schemes also allows us to precisely calibrate the atom number in the absorption images.

To model and fit the entire evolution of atom number $N_{\text{tot}}$ and filling $f$, we separate the atoms into two groups: those isolated in singly-occupied sites $N_1$ and those in multiply-occupied sites $N_2=N_{\text{tot}}-N_1$. Isolated atoms decay slowly with lifetime $t_1\gg$1~s. Radiative collisions cause atoms to leave a lattice site in a shorter time $t_2\ll 1$~s. To model the increase in filling $f$, we introduce the retention probability $\beta$ that the $N_2$ atoms will end up alone in a lattice site after LACs, which increase the number of isolated atoms $N_1$. The rest of the $N_2$ atoms are assumed lost with probability $1-\beta$. Our model provides a two-term exponential fit function to the data, from which we determine $t_1$, $t_2$ and $\beta$ ~\cite{supplement}. 

For the blue-, red-detuned, and near resonant OP data, we obtain $\beta = 54, 27, \text{and } 17\%$, respectively. The higher retention probability $\beta>50\%$ for blue-detuned OP indicates that atoms subject to LACs in a lattice site are more likely to hop and settle in singly-occupied sites. For red-detuned and near-resonant OP, on the other hand, the lower value of $\beta$ suggests that atoms have a higher probability to escape from the lattice, likely in the form of molecules \cite{burnett_laser-driven_1996}.

For the blue-detuned OP with $\Delta/2\pi = +9.9$~MHz, which yields the highest filling, the observed LAC dynamics are remarkably slow with timescales $t_2=39(4)$~ms for the increase in filling fraction and $49(11)$~ms for the initial atom decay. This should be compared with the light-induced collision rate of $\gamma_{\text{LAC}}=0.5$/ms which we estimate for two Cs atoms in a lattice site from the observed photon scattering rate \cite{supplement} with the theoretical model \cite{burnett_laser-driven_1996, vuletic_suppression_1999}. In the weak-field regime $\gamma_{\text{LAC}}\ll\Gamma$, where $\Gamma$ is the Cs excited state linewidth, scattering atoms are excited to the molecular potential once and stay for $1/\Gamma=30$ns. The excited state lifetime is much shorter than the time needed for atoms to fully slide down the repulsive potential $\tau=190$~ns, hence the radiative collision only imparts kinetic energy of $\hbar\Delta/(1+\Gamma^2\tau^2)=k_B\times11\mu$K to the atom pair \cite{supplement}.

Given the lattice trap depth $k_B\times150~\mu$K, it takes $\approx60$~ms for an atom to undergo $\approx30$ radiative collisions before gaining enough kinetic energy to hop to a neighboring site. In a dense sample prepared with $\bar{n}\approx2$ atoms per site, stochastic redistribution requires an average of $e^{\bar{n}}\approx7$ hopping events before an atom settles into a vacant site. This process continues until all atoms settle to a singly occupied site. We attribute the observed slow dynamics of the filling fraction and atom loss to these multi-LACs and multi-hopping processes. Ejection of atoms through multiple LACs events is also discussed in a recent tweezer experiment \cite{jenkins_ytterbium_2022}.

Compared with blue-detuned OP, dynamics with red-detuned and near-resonant beam are substantially faster with $t_2=11$ and 6~ms for the rise in filling fraction and $t_2=35$ and 25~ms for the total atom number loss, respectively. Here the shorter timescales, smaller gain in filling and greater atom loss can all be explained by the greater radiative collision loss \cite{burnett_laser-driven_1996}. The strong radiative loss with $\beta\ll$1 quickly depletes atoms in multiply-occupied sites, resulting in shorter LAC timescales for red-detuned and near resonant OP.

\begin{figure}[t]
\centering
\includegraphics[width=0.99\linewidth]{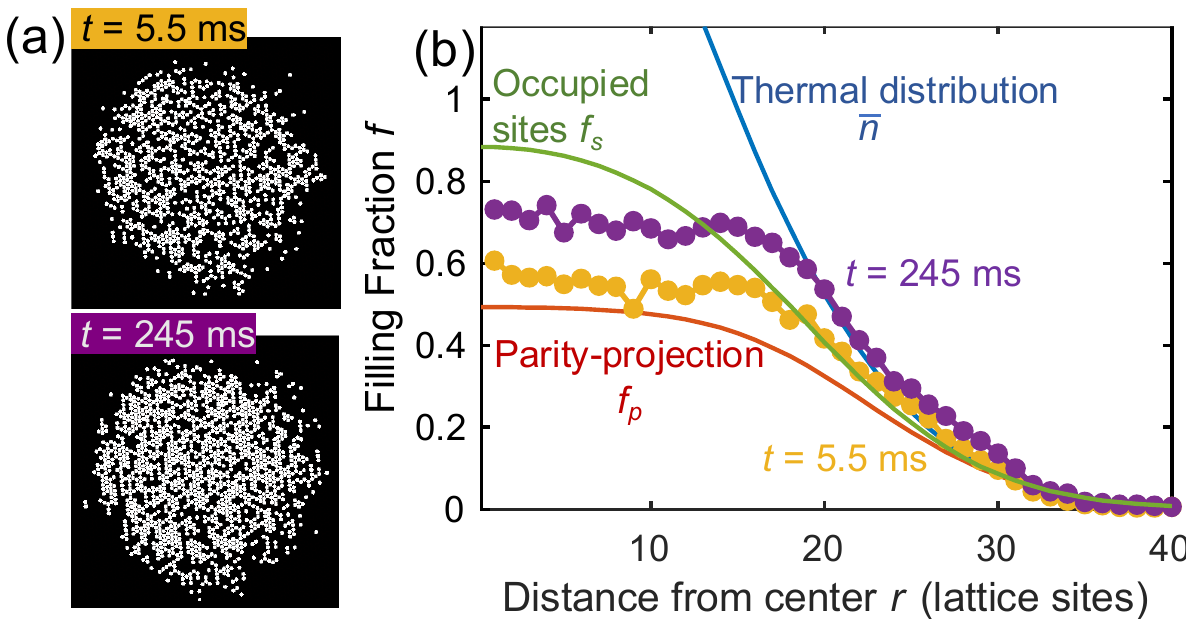} 
   \caption{Light-induced redistribution of atoms in the lattice by blue-detuned OP ($\Delta/2\pi=9.9~$MHz, $I= 1.2I_{\text{sat}}$). (a) Atomic distributions extracted from site-resolved fluorescence images with a short (left) and a long (right) LAC-cooling stage. (b) Filling fraction $f$ as a function of distance $r$ from the trap center at $t = 5.5$ (yellow) and $245$~ms (purple) based on 20 experimental shots. We compare the data with a Gaussian fit $\bar{n}$ (blue line) to the tail of the $t = 5.5$ ms data, the parity-projected distribution $f_p$ (red line), and the distribution of the initially occupied sites $f_s$ (green line).}
   \label{fig:spatmain}
\end{figure}

\subsection{Spatial redistribution of atoms}
Finally, we directly visualize the effect of atom redistribution in the lattice by comparing site-resolved images of atoms before and after the LAC-cooling process. Starting with a thermal sample of atoms loaded into the 2D lattice, we expect the initial average atom distribution $\bar{n}(r)$ to be Gaussian with lattice site occupancies $k$ following the Poisson distribution $P(k)$. We then apply the LAC-cooling with optimal OP parameters.

For a very short LAC-cooling time of $t=5.5$ ms, the site-resolved images reflect the parity-projected site distribution of the initial cloud. We observe a flat top distribution in the central region with filling fraction of $f \approx 0.55$ over a radius of 16 lattice sites surrounded by a Gaussian tail (Fig.~\ref{fig:spatmain}a). A fit to the tail determines the initial thermal distribution $\bar{n}$. Together with the total atom number of $N_{\text{tot}}(0)= 2000$ from absorption imaging, we determine the initial site occupancy $\bar{n}(r)=ae^{-r^2/2\sigma^2}$ where $a=2.2$ atoms per site is the site occupancy at the trap center and $\sigma = 12$ lattice sites is the width (blue curve in Fig.~\ref{fig:spatmain}). 
Our measurement of the spatial distribution at $t=5.5$~ms can be compared with ideal parity projection $f_p=\frac{1}{2}(1-e^{-2\bar{n}})$. Our measured central filling fraction is slightly higher than the expected value of $0.49$ likely due to the weak blue-detuned LACs induced during imaging. 

After the atomic distribution settles, we analyze images at $t = 245$~ms. Here, the spatial distribution shows a number of features. The filling within the central region with a radius of 16 lattice sites uniformly increases to $0.72$. Furthermore the filling fraction is uniformly higher than the initial distribution by $>20~\%$ over the entire sample, including the low-density tail where LACs are rare. Here the measured filling fraction in the tail even exceeds the initial probability for a site to be occupied $f_s=1-P(0)=1-e^{-\bar{n}}$ (Fig.~\ref{fig:spatmain}b). Exceeding this occupied-site limit in the low density tail suggests that a net flow of atoms migrates from the trap center to the edge, where empty sites are more abundant and radiative collisions are less likely. The migration of atoms to the wings is evidence of redistribution.

\section{Conclusion}
We expect that the light-induced redistribution is a generic radiative process on two or more atoms confined in a lattice site or optical tweezer.  Our observation of retaining about 50$\%$ of the atoms after many light-assisted collisions in a lattice site also suggests that the loss probability per radiative collisions can be negligible on the per-collision basis. We may then model the radiative collisions as an effective strong repulsion between atoms, which drives excess atoms toward vacant sites. These results suggest a platform for studying non-equilibrium many-body dynamics driven by light-assisted interactions in optical lattices. 

\textcolor{black}{In summary, we show that blue-detuned optical pumping can efficiently suppress radiative loss and redistribute atoms across lattice sites. For quantum information and simulation applications requiring unity filling, our scheme provides a fast and scalable route to high-filling atom arrays, reducing the number of tweezer rearrangement moves needed to eliminate residual vacancies. Looking ahead, the observed filling fractions of 70–80$\%$ can likely be improved further through more compact optical lattice geometries, optimized optical cooling, and optical pumping configurations that more efficiently recapture radiatively heated atoms.}

\section*{Acknowledgments}
We acknowledge J.~Trisnadi and M.~Zhang for their early work on the project. We would like to thank L. Khaykovich for helpful conversations on reabsorption. We acknowledge support from the National Science Foundation Quantum Leap Challenge Institute for Hybrid Quantum Architectures and Networks (HQAN) (NSF OMA-2016136) and the NSF Graduate Research Fellowship under Grant No. DGE 1746045.

\section*{Data Availability}
There are no publicly available research data or software supporting this manuscript. Requests for further information or data should be sent to the authors.

\bibliography{refs20260518}

\clearpage
\setcounter{section}{0}
\renewcommand{\thesection}{\Roman{section}}
\section*{Supplement to Achieving high filling of an optical lattice by light-assisted redistribution of atoms}
\subfile{supplement}

\end{document}

%% file: supplement.tex
\renewcommand{\thefigure}{S\arabic{figure}}
\setcounter{figure}{0}

\section{Experimental Setup and Hardware}\label{setup}
\begin{figure}[h]
\centering
\includegraphics[width=0.99\linewidth]{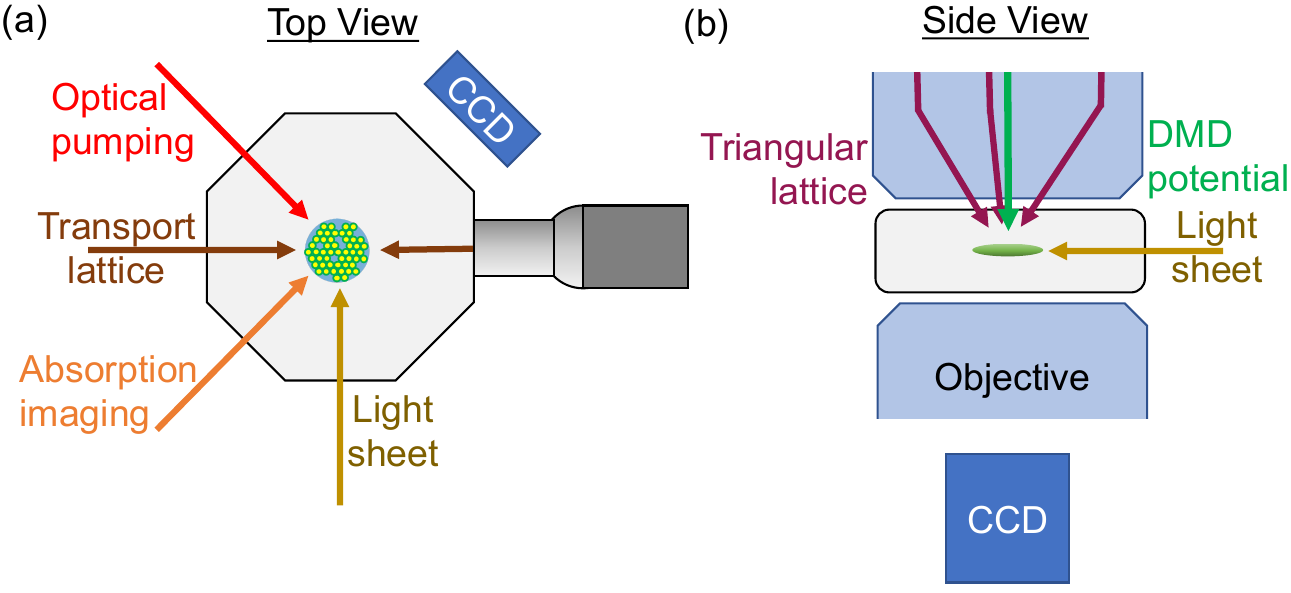}
\caption{Apparatus and optical beam configuration. a) The Raman sideband cooling optical pumping (OP) beam, absorption imaging beam, transport lattice, and light sheet dipole trap are in the xy plane. b) The triangular lattice and DMD potential are projected through the upper microscope objective (shown in blue).}
\label{fig:beams}
\end{figure}

The Quantum Matter Synthesizer apparatus consists of two chambers: a stainless steel chamber and a glass science cell \ref{fig:beams}. The glass science cell is sandwiched between two NA = 0.8 microscope objectives from Special Optics, one above and one below. The triangular optical lattice and the green DMD potential are both projected through the upper objective. Fluorescence from the atoms is collected through the lower microscope objective and imaged on an Andor Ikon-M CCD. The fluorescence imaging details can be found in Section \ref{imaging}. Absorption images are acquired using a Pixelfly CCD in the lattice plane. Details on calibrating the atom number from the absorption images can be found in Section \ref{numcal}. 

We use a digital micromirror device (DMD) to create custom green optical potentials that we project onto the atoms, as seen in Fig.~\ref{fig:green}). We illuminate a Texas Instruments DLP4500 digital micromirror device (DMD) with 5W of 532 nm laser light from a Lighthouse Photonics Sprout Solo laser. $55\%$ of the light is diffracted into the first order of the DMD and co-propagates with the optical lattice into the upper objective \cite{trisnadi_design_2022}. The DMD sits in an image plane of the optical system such that the pattern from the DMD is projected onto the atoms. Since the green light is blue-detuned from the Cs D2 line, it creates a repulsive potential  \cite{piotrowicz-two-dimensional-2013}.

The optical lattice is formed by three Gaussian beams that propagate through the upper objective and interfere. Each beam has a $1/e^2$ radius of 40 $\mu$m. Details of the beam geometry can be found in Ref.\cite{trisnadi_design_2022}. The lattice constant is $a=880~$nm, the trap depth at the center of the lattice is $U=k_B\times 150~\mu$K and the trap frequency is $\omega_r=2\pi\times75$~kHz. The trap depth varies from 150 to 140~$\mu$K within the central 217-site region of interest.

The 1064~nm light sheet dipole trap has trap depth 5.8~mK and trap frequency $\omega_z=2\pi\times$18~kHz in the vertical direction and $\omega_{r}=2\pi\times1.2$~kHz in the horizontal plane.

\begin{figure}[t]
\centering
\includegraphics[width=0.90\linewidth]{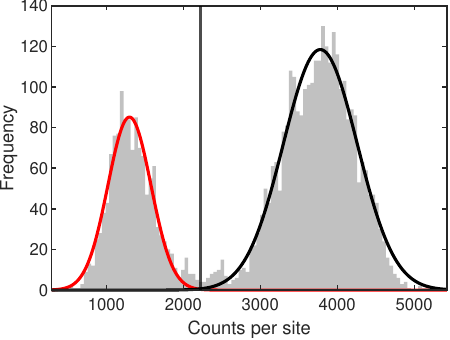}
\caption{Histogram of counts per lattice site in a 217-site region of interest in a 20-shot data set with 200 ms exposure time. We fit a Gaussian to the 0 peak (shown in red) and the 1 peak (shown in black) and set a threshold (black vertical line). The fidelity is 0.9994.}
\label{fig:hist}
\end{figure}

\section{Experimental Procedure}
\label{procedure:}
The experiment starts with $10^7$ Cs atoms in a magneto-optical trap in the stainless steel chamber, loaded in 3~s. After polarization gradient cooling and Raman sideband cooling in the stainless steel chamber, the atoms are transferred by an optical lattice conveyor belt to the glass cell science chamber over 450~ms. In the glass cell, atoms are further cooled to ~15 $\mu$K via degenerate Raman sideband cooling (dRSC) in a 2D triangular optical lattice for 55~ms. The glass cell beam geometry is pictured in Fig.~\ref{fig:beams}. Next, we ramp on the light sheet dipole trap that confines atoms to a single layer. We then take 4~ms to ramp off the triangular lattice, leave it off for 5~ms, then take another 4~ms to ramp it back to full power. This compression step allows atoms to collect in the center of the light sheet so there is more than one atom per site on average when the lattice turns back on. Once the lattice is back on, the OP and repump beams are applied and the OP beam and light sheet are modulated out of phase at 1 MHz to avoid light shifts \cite{hutzler_eliminating_2017}. This is the LAC-cooling stage where optimizing the OP intensity/frequency/duration induces high filling. After that, either absorption imaging is performed or we continue dRSC in the lattice and light sheet at higher OP power to induce fluorescence and take site-resolved images.

\section{Degenerate Raman sideband cooling}\label{cooling}
In the glass cell, we perform degenerate Raman sideband cooling (dRSC) \cite{vuletic_degenerate_1998,kerman_beyond_2000,han_3d_2000} in the 935~nm triangular optical lattice. The $\sigma^+$ optical pumping (OP) beam is near the $|6S_{1/2}, F=3\rangle$ to $|6P_{3/2}, F'=2\rangle$ transition. We employ a $\sigma^+$ repump beam on the $|6S_{1/2}, F=4\rangle$ to $|6P_{3/2}, F'=4\rangle$ transition which co-propagates with the OP beam. Raman coupling is provided by the 935~nm lattice. The bias magnetic field compensates for Earth's field and the field induced by the lattice beams to create a field in the plane of the lattice that sets the quantization axis along the OP beam direction and shifts the Zeeman sublevels of $|6S_{1/2}, F=3\rangle$ by an energy equal to $\hbar\omega$ where $\omega$ is the lattice trap frequency 2$\pi\times75$ kHz. Once we introduce the light sheet to confine atoms into a single layer, we modulate the light sheet and OP beam at 1 MHz during dRSC in the lattice to avoid light shifts \cite{hutzler_eliminating_2017}. The optimal cooling conditions for single layer cooling are OP beam intensity of 0.33$I_{\text{sat}}$ and repump beam intensity $\sim I_{\text{sat}}$.

\section{Site-resolved fluorescence imaging}
\label{imaging}
To perform site-resolved fluorescence imaging, we continue dRSC in the lattice, continuing to alternate the OP and light sheet. We increase the OP power to $16I_{\text{sat}}$ to make the atoms fluoresce more. We collect the fluorescence through the lower microscope objective and image with an Andor Ikon-M CCD with 200~ms exposure time.  A high spatial resolution of 800~nm is reached, consistent with the diffraction limit. We  obtain the number of counts per site by performing a Fourier transform then using a kernel deconvolution. We then create a histogram of the counts per site (see Fig.~\ref{fig:hist} for an example histogram), fit Gaussians to the 0-atom peak (red curve) and the 1-atom peak (black curve), then set a threshold that cuts off an equal area under each Gaussian to distinguish occupied from unoccupied sites. The fidelity is calculated from the overlap between the Gaussians \cite{trisnadi_design_2022}.

\section{Optical potential based on DMD}
We project a static green pattern that is calibrated for the green light to hit particular lattice sites. By using an acousto-optic modulator to modulate the intensity of the green light pattern at the trap frequency (75 kHz), we remove atoms from certain columns of sites in the lattice. We project a series of 5 stripes, each ~10 DMD pixels wide. These stripes are calibrated to shine along a lattice vector every 5 sites. The stripes remove atoms from 2 lattice columns. 

We begin with the usual sequence: an initial stage of Raman sideband cooling in the lattice only, followed by turning on the light sheet and performing light sheet compression (release atoms from lattice into light sheet then turn lattice back on). We then cool for 24 ms before turning on and modulating the green light potential at 75 kHz (the lattice trap frequency) for 50 ms. After that, we perform the LAC-cooling stage with the OP beam either near resonance ($\Delta/2\pi = 4.3$ MHz, $I < I_{\text{sat}}$) or blue-detuned ($\Delta/2\pi = 9.9$ MHz, $I \approx I_{\text{sat}}$) for 70 ms. Then we take a fluorescence image and observe whether atoms redistribute into the emptied sites. The preparation data is taken by turning on and modulating the green potential after 24 ms of cooling and keeping it on throughout the fluorescence imaging process. The filling fractions of the green and non-green areas can be found in Table \ref{table:green}.

\begin{table}[H]
\centering
\caption{Filling fraction results from Fig.~\ref{fig:green}}
\begin{tabular}{lccc}
\hline
\textbf{Conditions} & Green column filling & Other column filling  \\ \hline
Preparation              & 0.014(7)          & 0.396(9)          \\
Blue-detuned OP          & 0.102(7)          & 0.259(9)           \\
Near-resonant OP              & 0.037(4)          & 0.210(7)         \\ \hline
\label{table:green}
\end{tabular}
\end{table}

\section{Atom number calibration}
\label{numcal}

The site-resolved fluorescence imaging, described in Section \ref{imaging}, is subject to parity projection so we always detect 0 or 1 atom per site. The absorption imaging, performed using 852~nm light, is not site-resolved but is not subject to parity projection. We measure the optical depth (OD) of the sample and calculate the atom number $N_{\text{tot}}$. For consistent atom number calibration, we count all the atoms in the fluorescence images (see Fig.~\ref{fig:counting}) and use that number $N_{\text{site}}$ to appropriately scale the absorption image atom number $N_{\text{tot}}$, which is otherwise subject to systematics. We expand the region of interest to contain the entire cloud and count all of the atoms in the fluorescence image by setting the threshold to distinguish occupied (1) from unoccupied (0) sites separately for each hexagonal``ring" of the lattice. The threshold becomes smaller toward the edges as the atoms fluoresce less due to the spatial inhomogeneity of the lattice. At late times, all lattice sites have either 1 or 0 atom and the absorption atom number and site-resolved imaging atom number should agree. We use the late time data to calibrate the absorption image number to match the site-resolved fluorescence data by multiplying it by a scaling factor of 1.75.

At late times, single particle loss from singly-occupied sites (background loss) accounts for the reduction in total particle number as well as the reduction in filling fraction. Thus we expect the decay rates to be the same for the site-resolved fluorescence data (filling) and $N_{\text{tot}}$. The discrepancy in the decay rates of the filling fraction $f$ and the $N_{\text{tot}}$, seen in Fig.~\ref{fig:time}, arises from the differing regions of interest. When we count all the atoms (Fig.~\ref{fig:counting}), the decay rate of $N_{\text{site}}$ and $N_{\text{tot}}$ agree ($6$~s).

\section{Model of particle number evolution}
\label{model}
We present a model to describe the dynamics of the distribution of atoms in lattice sites in the entire system: atoms in singly-occupied sites $N_1$ and atoms in multiply-occupied sites $N_2$. \textcolor{black}{Notably, $N_2$ includes particles in sites with two, three, four or more atoms.} We model their dynamics with the following coupled differential equations:
\begin{align}
    \dot{N}_1 &= -\frac{N_1}{t_1} + \beta \frac{N_2}{t_2} \\
    \dot{N}_2 &= -\frac{N_2}{t_2},
\end{align}
where $t_1$ is the single particle loss time scale, $t_2$ is multiply-occupied site decay time scale and $\beta$ is the retention probability, defined as the probability that the loss of an atom from a multiply-occupied site results in a gain of an atom in a singly-occupied site.

This model describes the filling fraction and atom number dynamics. The analytic solutions for these populations are:
\begin{align}
    N_2(t) &= N_2(0) e^{-t/t_2} \label{eq:nums1}\\
    N_1(t) &= N_1(0) e^{-t/t_1} + \frac{\beta t_1 N_2(0)}{t_1 - t_2} \left( e^{-t/t_1} - e^{-t/t_2} \right)
    \label{eq:nums2}
\end{align}

\begin{figure*}[t]
\small
\centering
\includegraphics[width=0.9\textwidth]{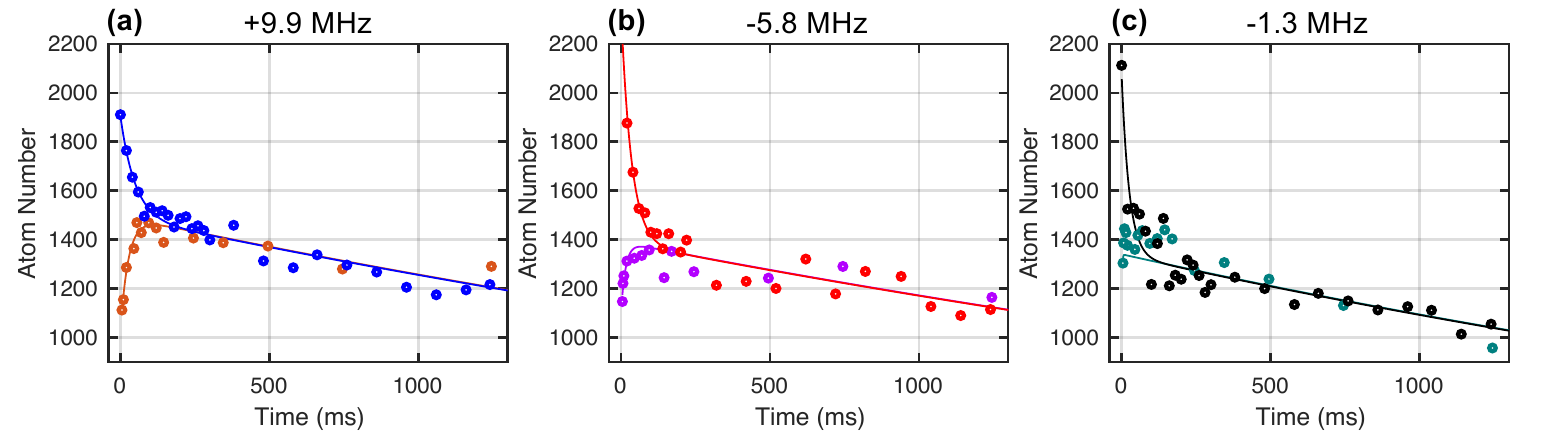}
\caption{Atom number evolution for (a) blue-detuned, (b) red-detuned, and (c) near-resonant optical pumping (OP). Site-resolved fluorescence data $N_{\text{site}}$ (orange, magenta, teal circles) and absorption data $N_{\text{tot}}$ (blue, red, black circles) are fit with two-term exponential functions (solid lines). Absorption data are scaled to match the tails of the total site-resolved fluorescence, which include all atoms rather than a 217-site ROI.}
\label{fig:counting}
\end{figure*}

\begin{table*}[t]
\small
\centering
\setlength{\tabcolsep}{4pt}
\renewcommand{\arraystretch}{1.1}
\begin{tabular}{cccccccccc}
\hline
 & \multicolumn{4}{c}{\textbf{Site-resolved imaging}} 
 & \multicolumn{4}{c}{\textbf{Absorption imaging}}
 & \textbf{Retention probability} \\
%\hline
\textbf{OP Detuning $\Delta/2\pi$} & $\Delta f$ & $t_1$ & $t_2$ & $\Delta N_1$
& $N_A$ & $N_B$ & $t_1$ & $t_2$  & $\beta$ \\
(MHz) &  & (s) & (ms) & & &  & (s) & (ms) &   \\
\hline
+9.9 & 0.174(5) & 8.7(7) & 39(4) & 480 & 1490(20) & 410(50) & 5.8(3) & 49(11)  & 54\% \\
-5.8 & 0.094(6) & 11(1)  & 11(3) & 360 & 1390(20) & 970(60) & 5.8(3) & 35(4)   & 27\% \\
-1.3 & 0.08(1)  & 2.3(2) & 6(4)  & 150 & 1340(20) & 750(60) & 4.9(3) & 25(5)  & 17\% \\
\hline
\end{tabular}
\caption{Fit results for filling fraction $f$ in central 217-site ROI (Fig.~\ref{fig:time}a), fit using Eq.~(\ref{eq:fillfit}). $f_0$ is constrained to 0.55. Fit results for total atom number $N_{\text{tot}}$ (Fig.~\ref{fig:time}b), fit using Eq.~(\ref{eq:numfit}). Fit results for site-resolved fluorescence image total atom number $N_{\text{site}}$, fit using Eq.~(\ref{eq:fillfit}). Right column is the retention probability $\beta$ calculated from Eq.~(\ref{eq:b}).}
\label{tab:combinedfits}
\end{table*}

We use these solutions to fit two sets of data: the site-resolved fluorescence images, from which we obtain the total parity-projected atom number $N_{\text{site}}$ (Fig.~\ref{fig:counting}), and the absorption images, from which we obtain the total atom number $N_{\text{tot}}$ (Fig.~\ref{fig:time}b and Fig.~\ref{fig:counting}).

We fit $N_{\text{site}}$, the parity-projected atom number shown in Fig.~\ref{fig:counting}, with a modified version of Eq.~(\ref{eq:nums2}), which describes the evolution of the number of atoms in singly-occupied sites. However, $N_{\text{site}}$ also includes all atoms in odd-occupied sites. As the system evolves, most of the gain in $N_{\text{site}}$ comes from the increase in singly-occupied sites $\Delta N_1$. We ignore any increase in triply- or quintuply- occupied sites since there are few of those and they rapidly decay. At time $t=0$, we include all atoms in odd-occupied sites in $N_{\text{site}}(0)$. Therefore we write the time dependence of $N_{\text{site}}(t)$ as follows:

\begin{equation}
    N_{\text{site}}(t) = N_{\text{site}}(0)e^{-t/t_1} + \Delta N_1(e^{-t/t_1}-e^{-t/t_2})
    \label{eq:siteatoms}
\end{equation}
where
\begin{equation}
    \Delta N_1 \approx \frac{\beta t_1 N_2(0)}{t_1 - t_2}
    \label{eq:siteatoms2}
\end{equation}
is the increase in the number of atoms in singly-occupied sites, based on Eq.~(\ref{eq:nums2}). We extract $\Delta N_1$ by fitting the site-resolved atom number versus time data with Eq.~(\ref{eq:siteatoms}); the results for the three different frequencies are given in Table \ref{tab:combinedfits}. 

We use this model to characterize the total atom number $N_{\text{tot}}$ by adding Eq.~(\ref{eq:nums1}) and Eq.~(\ref{eq:nums2}). We fit the resulting 2-term exponential to the total atom number $N_{\text{tot}}$ from the absorption imaging, shown in Fig.~\ref{fig:counting} and Fig.~\ref{fig:time}b:
\begin{align}
    N_{\text{tot}}(t) &= N_1(t) + N_2(t) \\
    &= N_Ae^{-t/t_{1}} +N_B e^{-t/t_{2}}
    \label{eq:numfit}
\end{align}

We extract $N_A$ and $N_B$ from the fits. $N_A + N_B$ is the initial particle number. $N_B$ describes how many particles are lost in the initial density-dependent loss stage while $N_A$ describes how many particles are present after the initial density-dependent loss stage.  Using the full expressions for $N_1(t)$ and $N_2(t)$ from Eqs.~\ref{eq:nums1} and \ref{eq:nums2},  we find that:
\begin{align} \label{eq:na}
    N_A &= N_1(0)+\frac{\beta t_1 N_2(0)}{t_1-t_2} = N_1(0)+ \Delta N_1\\
    N_B &= N_2(0)-\frac{\beta t_1 N_2(0)}{t_1-t_2}=N_2(0)-\Delta N_1
    \label{eq:nb}
\end{align}

To calculate the retention probability $\beta$, we derive $\Delta N_1$ from the fit to the total site-resolved atom number data $N_{\text{site}}$ (Eq.~(\ref{eq:siteatoms})) and $t_1$, $t_2$, and $N_B$ from the fit to the total atom number data $N_{\text{tot}}$ (Eq.~(\ref{eq:numfit})). 
The retention probability $\beta$ is calculated using Eqs.~\ref{eq:siteatoms2} and \ref{eq:nb}:
\begin{equation}
    \beta = \frac{\Delta N_1 (t_1-t_2)}{t_1 (N_B + \Delta N_1)}
    \label{eq:b}
\end{equation}

For the blue-detuned data, we determine the following parameters: $\Delta N_1 = 480$ atoms, $t_2 = 49 \text{ ms}$, $t_1 = 5.8 \text{ s}$ and $N_B = 410$ atoms.
From the blue-detuned OP data, the retention probability is $\beta=0.54$. We similarly calculate $\beta$ for the red-detuned and resonant OP data and find them to be 0.27 and 0.17, respectively.

The filling fraction data (Fig.~\ref{fig:time}a) characterizes the fraction of filled sites in a 217-site central ROI over time. The time dependence approximately follows Eq.\ref{eq:nums2}, the evolution of singly occupied sites, since the gain in filling is mainly from the gain in singly-occupied sites and the slow loss comes from single particle decay.

We write:
\begin{equation}
    f(t) = f_0e^{-t/t_1} + \Delta f(e^{-t/t_1}-e^{-t/t_2})
    \label{eq:fillfit}
\end{equation}

where $f_0$ is the initial filling and $\Delta f$ is the increase in filling fraction. We use Eq.~(\ref{eq:fillfit}) to fit the filling versus time data in Fig.~\ref{fig:time}a. We constrain the initial filling $f_{0}$ to 0.55 since these experiments all share the same conditions up until $t = 0$, when the LAC-cooling stage begins.

\section{Radiative heating and collision rate}
\label{collision}

The long-range molecular potential between one Cs atom in the ground state and one Cs atom in the excited state is given by $V(r) = \frac{C_3}{r^3}$. For the strongest molecular line, 2g, $C_3 =  f_3 \hbar \Gamma (\frac{\lambda}{2\pi})^3=6.6 \times 10^{-48} \text{ J m}^3$ \cite{burnett_laser-driven_1996, vuletic_suppression_1999} where $\Gamma = 2 \pi \times 5.3$ MHz is the Cs natural linewidth, $f_3 =0.75$ is the dipole-dipole coefficient, and $\lambda = 852$ nm is the wavelength. The Condon point $R_c$, defined as the separation at which two atoms absorb a photon at detuning $\Delta$ is given by $R_c = \left( \frac{C_3}{\hbar \Delta} \right)^{1/3}$. 

\begin{figure}[b]
\small
\centering
\includegraphics[width=0.5\textwidth]{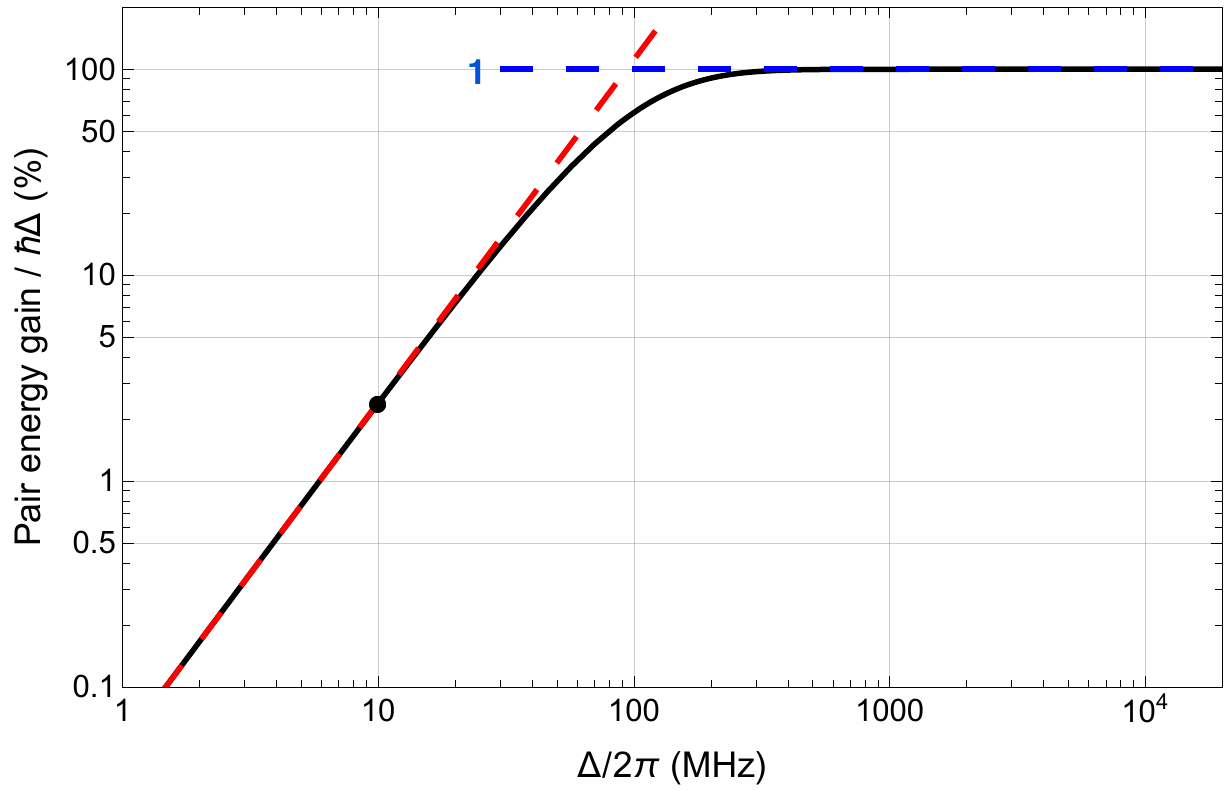}
\caption{Atom pair energy gain from a single repulsive LAC as a function of OP frequency detuning. The blue dashed line describes the asymptotic behavior at large OP frequency detunings, where the atom pair gains the entirety of the OP detuning energy $\hbar\Delta$, while the red dashed line describes the small OP detuning limit where the atom pair gains a fraction of that energy given by $5.5\times10^{-4} (\Delta/2\pi \text{ MHz})^{5/3}$. The dot indicates our detuning $\Delta/2\pi = 10$~MHz.}

\label{fig:energygain}
\end{figure}

% more general about energy vs detuning
Upon excitation by the OP beam, the molecular potential repels the atoms with a relative acceleration on the order of $a = \frac{6C_3}{m R_c^4}$, where $m$ is the Cs mass, until the pair either decays to the ground state or the kinetic energy reaches $\hbar \Delta$. The time scale for the atom pair to gain the full kinetic energy transferred from the molecular potential is given by $\tau = \sqrt{\frac{4 \hbar \Delta}{m a^2}}$. Given a limited  excited state lifetime $1/\Gamma$ the actual energy gained by the pair is limited to $\frac{\hbar\Delta}{1+\Gamma^2\tau^2}$, see Fig.~\ref{fig:energygain}.

% more specific to our parameters
At detuning $\Delta=2\pi \times 10$ MHz, we obtain the Condon point $R_c = 100$~nm for the $2g$ molecular potential. For the Cs D2 line, the excited state lifetime $1/\Gamma$ = 30~ns is short compared to the energy transfer time scale $\tau=190$~ns. Thus the atom pair only gains a small fraction of the total detuning, $0.024\Delta = 2\pi \times 240$~kHz per LAC, or approximately $5.7\mu$K per atom.

At detuning $\Delta=2\pi \times 10$ MHz, the collision rate is
\begin{equation}
    \gamma_{\text{LAC}}=4\pi^2b_c^2f_3(\lambda/2\pi)^3h_c\Gamma_{\text{atom}}\rho
\end{equation}
where $b_c=0.5$ is the unitless coupling factor for the 2g molecular state which has the strongest radiative coupling to ground state atoms, $h_c=0.016$, $f_3=0.75$ for the 2g molecular state, and $\Gamma_{\text{atom}}=1.4\times10^4$/s is the single atom photon scattering rate extracted from experiment, and $\rho$ is the atom density \cite{vuletic_suppression_1999}. For $N=2$ atoms in a lattice site with trap frequencies $(\omega_x, \omega_y, \omega_z) = 2\pi \times (75, 75, 18)$ kHz and temperature $T = 15 \mu$K, the density is approximated as
\begin{equation}
    \rho = \frac{N\omega_x \omega_y \omega_z}{(2 \pi)^{3/2}} ( \frac{m}{k_B T} )^{3/2}= 1.1\times 10^{20} ~\text{m}^{-3}.
\end{equation}

We extract $\Gamma_{\text{atom}}$ from the number of CCD counts per atom in our site-resolved fluorescence images. From our images, occupied sites have on average $c=1.9\times10^4$ counts/s (Fig.~\ref{fig:hist}). The imaging single atom scattering rate $\Gamma_{\text{image}}=\frac{cs}{q_e e_c}=9.5\times10^4$/s, where $s=0.9$ $e^-$ per CCD count is the sensitivity, $q_e=0.9$ is the quantum efficiency of the CCD (Andor Ikon-M) at 852 nm, and $e_c=20\%$ is the collection efficiency of the imaging system with N.A. = 0.8. The imaging is performed at high intensity $I=16I_{\text{sat}}$ whereas we are interested in finding $\Gamma_{\text{atom}}$ at the LAC-cooling intensity from Fig.~\ref{fig:time}, $I=1.2I_{\text{sat}}$. To scale $\Gamma_{\text{atom}}$ to the lower OP intensity, we calculate the single atom photon scattering rate from theory using the typical equation 
\begin{equation}
    \Gamma_{\text{theory}}= \frac{\Gamma}{2}\frac{I/I_{\text{sat}}}{1+I/I_{\text{sat}}+4\Delta^2/\Gamma^2}
\end{equation}
\cite{jooya_photon_2013}, where $I_{\text{sat}}$ is the saturation intensity, $\Gamma=2\pi\times5.3$ MHz is the natural linewidth, and $\Delta=2\pi\times 10$ MHz is the optical pumping detuning. For $I=16I_{\text{sat}}$, the theoretical photon scattering rate per atom $\Gamma_{\text{theory}}=8.5 \times 10^6$/s whereas for $I=1.2I_{\text{sat}}$, it is $1.2\times10^6$/s. We then divide the experimentally derived $\Gamma_{\text{image}}=9.5\times10^4$/s by the ratio between the theoretical values at the two intensities to obtain an estimated $\Gamma_{\text{atom}}=1.4\times 10^4$/s. The order-of-magnitude difference between $\Gamma_{\text{atom}}$ and $\Gamma_{\text{theory}}$ comes from the ongoing Raman sideband cooling; the atoms spend much of their time in the dark state and thus scatter fewer photons.

This model yields a calculated radiative collision rate $\gamma_{\text{LAC}}=0.5$/ms for two atoms in a lattice site. Given the lattice trap depth of $k_B\times150 ~\mu$K $= h\times 3.13$ MHz, we estimate that the cumulative energy gain requires 30 radiative collisions and 60 ms for an atom to leave the lattice site. \textcolor{black}{In the case of blue-detuned optical pumping, in particular, each radiative collision has a probability $p\approx 1$ of keeping the atoms in the atomic state and $1-p\ll1$ of losing them into deeply-bound molecules. We may write the measured retention probability as $\beta\approx p^{30}=0.54$, which suggests a high retention probability $p=98\%$ per radiative collision. This is consistent in order of magnitude with the prediction from Ref.~\cite{burnett_laser-driven_1996} that radiative collision loss rate with blue-detuned light can be a factor of $10^2$ to $10^3$ lower than the resonant photoassociation rate.}

%% file: refs20260518.bib
@article{jooya_photon_2013,
  author  = {Jooya, Kais and Musterer, Nam and Madison, Kirk W. and Booth, James L.},
  title   = {Photon-scattering-rate measurement of atoms},
  journal = {Physical Review A},
  volume  = {88},
  pages   = {063401},
  year    = {2013}
}

@article{pampel_quantifying_2025_1,
  author   = {Pampel, S. K. and Marinelli, M. and Brown, M. O. and D'Incao, J. P. and Regal, C. A.},
  title    = {Quantifying Light-Assisted Collisions in Optical Tweezers across the Hyperfine Spectrum},
  journal  = {Physical Review Letters},
  volume   = {134},
  pages    = {013202},
  year     = {2025},
  doi      = {10.1103/PhysRevLett.134.013202}
}

@misc{supplement,
  note    = {{See Supplemental Material at the end of this document for experimental details, atom number versus time model, and collision rate calculation, which includes Refs. [43-46].}}
}

@article{piotrowicz-two-dimensional-2013,
	author = {M. J. Piotrowicz and M. Lichtman and K. Maller and G. Li and S. Zhang and L. Isenhower and M. Saffman},
	journal = {Physical Review A},
	pages = {013420},
	title = {Two-dimensional lattice of blue-detuned atom traps using a projected Gaussian beam array},
	volume = {88},
	year = {2013}}

@article{bakr_quantum-gas_2009,
	author = {Waseem S. Bakr and Jonathon I. Gillen and Amy Peng and Simon F{\"o}lling and Markus Greiner},
	journal = {Nature},
	pages = {74-77},
	title = {A quantum gas microscope for detecting single atoms in a {H}ubbard-regime optical lattice},
	volume = {462},
	year = {2009}}

@article{singh_dual-element_2022,
  author   = {Singh, K. and Anand, S. and Pocklington, A. and Kemp, J. T. and Bernien, H.},
  title    = {Dual-Element, Two-Dimensional Atom Array with Continuous-Mode Operation},
  journal  = {Physical Review X},
  volume   = {12},
  pages    = {011040},
  year     = {2022}
}

@article{barnes_assembly_2022,
  author   = {Barnes, K. and Battaglino, P. and Bloom, B. J. and Cassella, K. and Coxe, R. and Crisosto, N. and King, J. P. and Kondov, S. S. and Kotru, S. and Larsen, S. C. and Lauigan, J. and Lester, B. J. and McDonald, M. and Megidish, E. and Narayanaswami, S. and Nishiguchi, C. and Notermans, R. and Peng, L. S. and Ryou, A. and Wu, T.-Y. and Yarwood, M.},
  title    = {Assembly and coherent control of a register of nuclear spin qubits},
  journal  = {Nature Communications},
  volume   = {13},
  pages    = {2779},
  year     = {2022}
}

@article{pichard_rearrangement_2024,
  author   = {Pichard, G. and Lim, \'{E}. and Bloch, J. and Vaneecloo, J. and Bourachot, L. and Both, G.-J. and M\'{e}riaux, G. and Dutartre, S. and Hostein, R. and Paris, J. and Ximenez, B. and Signoles, A. and Browaeys, A. and Lahaye, T. and Dreon, D.},
  title    = {Rearrangement of individual atoms in a 2000-site optical-tweezer array at cryogenic temperatures},
  journal  = {Physical Review Applied},
  volume   = {22},
  pages    = {024073},
  year     = {2024}
}

@article{barredo_atom-by-atom_2016,
  author   = {Barredo, D. and de L\'{e}s\'{e}leuc, S. and Lienhard, V. and Lahaye, T. and Browaeys, A.},
  title    = {An atom-by-atom assembler of defect-free arbitrary two-dimensional atomic arrays},
  journal  = {Science},
  volume   = {354},
  pages    = {1021},
  year     = {2016}
}

@article{manetsch_tweezer_2025,
  author   = {Manetsch, H. J. and Nomura, G. and Bataille, E.  and Lv, X. and Leung, K. H. and Endres, M.},
  title    = {A tweezer array with 6100 highly coherent atomic qubits},
  journal  = {Nature},
  volume   = {647},
  pages    = {60},
  year     = {2025}
}

@article{bluvstein_logical_2024,
  author   = {Bluvstein, D. and Evered, S. J. and Geim, A. A. and Li, S. H. and Zhou, H. and Manovitz, T. and Ebadi, S. and Cain, M. and Kalinowski, M. and Hangleiter, D. and Bonilla Ataides, J. P. and Maskara, N. and Cong, I. and Gao, X. and Sales Rodriguez, P. and Karolyshyn, T. and Semeghini, G. and Gullans, M. J. and Greiner, M. and Vuleti\'{c}, V. and Lukin, M. D.},
  title    = {Logical quantum processor based on reconfigurable atom arrays},
  journal  = {Nature},
  volume   = {626},
  pages    = {58},
  year     = {2024}
}

@article{sherson_single-atom-resolved_2010,
  author   = {Sherson, J. F. and Weitenberg, C. and Endres, M. and Cheneau, M. and Bloch, I. and Kuhr, S.},
  title    = {Single-atom-resolved fluorescence imaging of an atomic {M}ott insulator},
  journal  = {Nature},
  volume   = {467},
  pages    = {68},
  year     = {2010}
}

@article{parsons_site-resolved_2015,
  author   = {Parsons, M. F. and Huber, F. and Mazurenko, A. and Chiu, C. S. and Setiawan, W. and Wooley-Brown, K. and Blatt, S. and Greiner, M.},
  title    = {Site-Resolved Imaging of Fermionic $^{6}\mathrm{Li}$ in an Optical Lattice},
  journal  = {Physical Review Letters},
  volume   = {114},
  pages    = {213002},
  year     = {2015}
}

@article{fuhrmanek_study_2012,
  author   = {Fuhrmanek, A. and Bourgain, R. and Sortais, Y. R. P. and Browaeys, A.},
  title    = {Study of light-assisted collisions between a few cold atoms in a microscopic dipole trap},
  journal  = {Physical Review A},
  volume   = {85},
  pages    = {062708},
  year     = {2012}
}

@article{depue_unity_1999,
  author   = {DePue, M. T. and McCormick, C. and Winoto, S. L. and Oliver, S. and Weiss, D. S.},
  title    = {Unity Occupation of Sites in a 3{D} Optical Lattice},
  journal  = {Physical Review Letters},
  volume   = {82},
  pages    = {2262},
  year     = {1999}
}

@article{buob_strontium_2024,
  author   = {Buob, S. and H\"{o}schele, J. and Makhalov, V. and Rubio-Abadal, A. and Tarruell, L.},
  title    = {A Strontium Quantum-Gas Microscope},
  journal  = {PRX Quantum},
  volume   = {5},
  pages    = {020316},
  year     = {2024}
}

@article{endres_single-site-_2013,
  author   = {Endres, M. and Cheneau, M. and Fukuhara, T. and Weitenberg, C. and Schau\ss, P. and Gross, C. and Mazza, L. and Ba\~{n}uls, M. C. and Pollet, L. and Bloch, I. and Kuhr, S.},
  title    = {Single-site- and single-atom-resolved measurement of correlation functions},
  journal  = {Applied Physics B},
  volume   = {113},
  pages    = {27},
  year     = {2013}
}

@article{su_fast_2025,
  author   = {Su, L. and Douglas, A. and Szurek, M. and H\'{e}bert, A. H. and Krahn, A. and Groth, R. and Phelps, G. A. and Markovi\'{c}, O. and Greiner, M.},
  title    = {Fast single atom imaging for optical lattice arrays},
  journal  = {Nature Communications},
  volume   = {16},
  pages    = {1017},
  year     = {2025}
}

@article{falcon_microsecond-scale_2025,
  author   = {Muzi Falconi, A. and Panza, R. and Sbernardori, S. and Forti, R. and Klemt, R. and Abdel Karim, O. and Marinelli, M. and Scazza, F.},
  title    = {Microsecond-scale high-survival and number-resolved detection of ytterbium atom arrays},
  journal  = {Physical Review Letters},
  volume   = {135},
  pages = {203402},
  year     = {2025}
}

@article{jaksch_cold_1998,
  author   = {Jaksch, D. and Bruder, C. and Cirac, J. I. and Gardiner, C. W. and Zoller, P.},
  title    = {Cold bosonic atoms in optical lattices},
  journal  = {Physical Review Letters},
  volume   = {81},
  pages    = {3108},
  year     = {1998}
}

@article{greiner_quantum_2002,
  author   = {Greiner, M. and Mandel, O. and Esslinger, T. and H\"{a}nsch, T. W. and Bloch, I.},
  title    = {Quantum phase transition from a superfluid to a {M}ott insulator in a gas of ultracold atoms},
  journal  = {Nature},
  volume   = {415},
  pages    = {39},
  year     = {2002}
}

@article{bakr_probing_2010,
  author   = {Bakr, W. S. and Peng, A. and Tai, M. E. and Ma, R. and Simon, J. and Gillen, J. I. and F\"{o}lling, S. and Pollet, L. and Greiner, M.},
  title    = {Probing the Superfluid--to--{M}ott Insulator Transition at the Single-Atom Level},
  journal  = {Science},
  volume   = {329},
  pages    = {547},
  year     = {2010}
}

@article{tomita_observation_2017,
  author   = {Tomita, T. and Nakajima, S. and Danshita, I. and Takasu, Y. and Takahashi, Y.},
  title    = {Observation of the {M}ott insulator to superfluid crossover of a driven-dissipative {B}ose-{H}ubbard system},
  journal  = {Science Advances},
  volume   = {3},
  pages    = {e1701513},
  year     = {2017}
}

@misc{eckner_assembling_2025,
  author   = {Eckner, W. J. and Yelin, T. L. and Cao, A. and Young, A. W. and Oppong, N. D. and Pollet, L. and Kaufman, A. M.},
  title    = {Assembling a {B}ose-{H}ubbard superfluid from tweezer-controlled single atoms},
  year     = {2025},
  note     = {arXiv:2512.24374}
}

@article{gyger_continuous_2024,
  author   = {Gyger, F. and Ammenwerth, M. and Tao, R. and Timme, H. and Snigirev, S. and Bloch, I. and Zeiher, J.},
  title    = {Continuous operation of large-scale atom arrays in optical lattices},
  journal  = {Physical Review Research},
  volume   = {6},
  pages    = {033104},
  year     = {2024}
}

@article{endres_atom-by-atom_2016,
  author   = {Endres, M. and Bernien, H. and Keesling, A. and Levine, H. and Anschuetz, E. R. and Krajenbrink, A. and Senko, C. and Vuletic, V. and Greiner, M. and Lukin, M. D.},
  title    = {Atom-by-atom assembly of defect-free one-dimensional cold atom arrays},
  journal  = {Science},
  volume   = {354},
  pages    = {1024},
  year     = {2016}
}

@article{grunzweig_near-deterministic_2010,
  author   = {Gr\"{u}nzweig, T. and Hilliard, A. and McGovern, M. and Andersen, M. F.},
  title    = {Near-deterministic preparation of a single atom in an optical microtrap},
  journal  = {Nature Physics},
  volume   = {6},
  pages    = {951},
  year     = {2010}
}

@article{carpentier_preparation_2013,
  author   = {Carpentier, A. V. and Fung, Y. H. and Sompet, P. and Hilliard, A. J. and Walker, T. G. and Andersen, M. F.},
  title    = {Preparation of a single atom in an optical microtrap},
  journal  = {Laser Physics Letters},
  volume   = {10},
  pages    = {125501},
  year     = {2013}
}

@article{brown_gray-molasses_2019,
  author   = {Brown, M. and Thiele, C. and Kiehl, C. and Hsu, T.-W. and Regal, C.},
  title    = {Gray-Molasses Optical-Tweezer Loading: Controlling Collisions for Scaling Atom-Array Assembly},
  journal  = {Physical Review X},
  volume   = {9},
  pages    = {011057},
  year     = {2019}
}

@article{jenkins_ytterbium_2022,
  author   = {Jenkins, A. and Lis, J. W. and Senoo, A. and McGrew, W. F. and Kaufman, A. M.},
  title    = {Ytterbium Nuclear-Spin Qubits in an Optical Tweezer Array},
  journal  = {Physical Review X},
  volume   = {12},
  pages    = {021027},
  year     = {2022}
}

@article{zhu_high-efficiency_2025,
  title = {High-Efficiency Loading of 2400 Ytterbium Atoms in Optical Tweezer Arrays},
  author = {Zhu, Jiawen and Chen, Changfeng and Zhou, Li and Xie, Xiangru and Jiang, Chenyang and Ding, Zhuoli and Wu, Fan and Yang, Fan and Wang, Guoqing and Gong, Qihuang and Zhang, Peng and Zhang, Sheng and Peng, Pai},
  journal = {Phys. Rev. Lett.},
  volume = {137},
  issue = {6},
  pages = {063201},
  numpages = {8},
  year = {2026},
  month = {Aug},
  publisher = {American Physical Society},
  doi = {10.1103/bp6k-8zmd},
  url = {https://link.aps.org/doi/10.1103/bp6k-8zmd}
}

@article{lester_rapid_2015,
  author   = {Lester, B. J. and Luick, N. and Kaufman, A. M. and Reynolds, C. M. and Regal, C. A.},
  title    = {Rapid production of uniformly-filled arrays of neutral atoms},
  journal  = {Physical Review Letters},
  volume   = {115},
  pages    = {073003},
  year     = {2015}
}

@article{trisnadi_design_2022,
  author   = {Trisnadi, J. and Zhang, M. and Weiss, L. and Chin, C.},
  title    = {Design and construction of a quantum matter synthesizer},
  journal  = {Review of Scientific Instruments},
  volume   = {93},
  pages    = {083203},
  year     = {2022}
}

@article{vuletic_degenerate_1998,
  author   = {Vuleti\'{c}, V. and Chin, C. and Kerman, A. J. and Chu, S.},
  title    = {Degenerate {R}aman Sideband Cooling of Trapped Cesium Atoms at Very High Atomic Densities},
  journal  = {Physical Review Letters},
  volume   = {81},
  pages    = {5768},
  year     = {1998}
}

@article{kerman_beyond_2000,
  author   = {Kerman, A. J. and Vuleti\'{c}, V. and Chin, C. and Chu, S.},
  title    = {Beyond Optical Molasses: 3{D} {R}aman Sideband Cooling of Atomic Cesium to High Phase-Space Density},
  journal  = {Physical Review Letters},
  volume   = {84},
  pages    = {439},
  year     = {2000}
}

@article{han_3d_2000,
  author   = {Han, D.-J. and Wolf, S. and Oliver, S. and McCormick, C. and DePue, M. T. and Weiss, D. S.},
  title    = {3{D} {R}aman Sideband Cooling of Cesium Atoms at High Density},
  journal  = {Physical Review Letters},
  volume   = {85},
  pages    = {724},
  year     = {2000}
}

@article{qiu_loading_2025,
  title = {Loading and Imaging Atom Arrays via Electromagnetically Induced Transparency},
  author = {Qiu, Emily H. and {\ifmmode \check{S}\else \v{S}\fi{}}umarac, Tamara and Niu, Peiran and Tsesses, Shai and Wassaf, Fadi and Spierings, David C. and Chen, Meng-Wei and Uysal, Mehmet T. and Bartlett, Audrey and Menssen, Adrian J. and Lukin, Mikhail D. and Vuleti{\ifmmode \acute{c}\else \'{c}\fi{}}, Vladan},
  journal = {PRX Quantum},
  volume = {7},
  issue = {2},
  pages = {020358},
  numpages = {13},
  year = {2026},
  month = {Jun},
  publisher = {American Physical Society},
  doi = {10.1103/fd15-4g1n},
  url = {https://link.aps.org/doi/10.1103/fd15-4g1n}
}

@article{khaykovich_saturation_1999,
  author   = {Khaykovich, L. and Friedman, N. and Davidson, N.},
  title    = {Saturation of the weak probe amplification in a strongly driven cold and dense atomic cloud},
  journal  = {The European Physical Journal D},
  volume   = {7},
  pages    = {467},
  year     = {1999}
}

@article{castin_reabsorption_1998,
  author   = {Castin, Y. and Cirac, J. I. and Lewenstein, M.},
  title    = {Reabsorption of Light by Trapped Atoms},
  journal  = {Physical Review Letters},
  volume   = {80},
  pages    = {5305},
  year     = {1998}
}

@article{gisbert_stochastic_2019,
  author   = {Gisbert, A. T. and Piovella, N. and Bachelard, R.},
  title    = {Stochastic heating and self-induced cooling in optically bound pairs of atoms},
  journal  = {Physical Review A},
  volume   = {99},
  pages    = {013619},
  year     = {2019}
}

@article{fung_efficient_2015,
  author   = {Fung, Y. H. and Andersen, M. F.},
  title    = {Efficient collisional blockade loading of a single atom into a tight microtrap},
  journal  = {New Journal of Physics},
  volume   = {17},
  pages    = {073011},
  year     = {2015}
}

@article{schlosser_collisional_2002,
  author   = {Schlosser, N. and Reymond, G. and Grangier, P.},
  title    = {Collisional Blockade in Microscopic Optical Dipole Traps},
  journal  = {Physical Review Letters},
  volume   = {89},
  pages    = {023005},
  year     = {2002}
}

@article{forster_number-triggered_2006,
  author   = {F{\"o}rster, L. and Alt, W. and Dotsenko, I. and Khudaverdyan, M. and Meschede, D. and Miroshnychenko, Y. and Reick, S. and Rauschenbeutel, A.},
  title    = {Number-triggered loading and collisional redistribution of neutral atoms in a standing wave dipole trap},
  journal  = {New Journal of Physics},
  volume   = {8},
  pages    = {259},
  year     = {2006}
}

@article{burnett_laser-driven_1996,
  author   = {Burnett, K. and Julienne, P. S. and Suominen, K.-A.},
  title    = {Laser-Driven Collisions between Atoms in a {B}ose-{E}instein Condensed Gas},
  journal  = {Physical Review Letters},
  volume   = {77},
  pages    = {1416},
  year     = {1996}
}

@article{vuletic_suppression_1999,
  author   = {Vuleti\'{c}, V. and Chin, C. and Kerman, A. J. and Chu, S.},
  title    = {Suppression of Atomic Radiative Collisions by Tuning the Ground State Scattering Length},
  journal  = {Physical Review Letters},
  volume   = {83},
  pages    = {943},
  year     = {1999}
}

@article{hutzler_eliminating_2017,
  author   = {Hutzler, N. R. and Liu, L. R. and Yu, Y. and Ni, K.-K.},
  title    = {Eliminating light shifts for single atom trapping},
  journal  = {New Journal of Physics},
  volume   = {19},
  pages    = {023007},
  year     = {2017}
}
